\documentclass[prb,twocolumn,showpacs,preprintnumbers,amsmath]{revtex4}
\usepackage{dcolumn}
\usepackage{bm}
\usepackage{graphicx}
\begin{document}

\title{Fishtail effect and vortex dynamics in LiFeAs single crystals}

\author{A. K. Pramanik}\affiliation{Institute for Solid State Research, IFW Dresden, D-01171 Dresden, Germany}
\author{L. Harnagea}\affiliation{Institute for Solid State Research, IFW Dresden, D-01171 Dresden, Germany}
\author{C. Nacke}\affiliation{Institute for Solid State Research, IFW Dresden, D-01171 Dresden, Germany}
\author{A. U. B. Wolter}\affiliation{Institute for Solid State Research, IFW Dresden, D-01171 Dresden, Germany}
\author{S. Wurmehl}\affiliation{Institute for Solid State Research, IFW Dresden, D-01171 Dresden, Germany}
\author{V. Kataev}\affiliation{Institute for Solid State Research, IFW Dresden, D-01171 Dresden, Germany}
\author{B. B\"{u}chner}\affiliation{Institute for Solid State Research, IFW Dresden, D-01171 Dresden, Germany}

\begin{abstract}
We investigate the fishtail effect, critical current density ($J_c$) and
vortex dynamics in LiFeAs single crystals. The sample exhibits a
second peak (SP) in the magnetization loop only with the field $||$ $c$-axis.
We calculate a reasonably high $J_c$,
however, values are lower than in `Ba-122' and `1111'-type FeAs-compounds.
Magnetic relaxation data imply a strong pinning which appears not to be due to
conventional defects. Instead, its behavior is similar
to that of the triplet superconductor Sr$_2$RuO$_4$. Our data
suggest that the origin of the SP may be related to a vortex
lattice phase transition. We have constructed the vortex phase
diagram for LiFeAs on the field-temperature plane.
\end{abstract}

\maketitle

The fishtail effect or the anomalous second peak (SP) in field
($H$) dependent magnetization ($M$) loops has been a subject of
intense research topic in the field of superconductors with both
low and high transition temperatures
($T_C$).\cite{roy,daeumling,klein,yoshizaki,kopylov,bugo,elbaum,rosenstein,khaykovich,ertas,giller,radzyner,abulafia}
This phenomenon is realized with the enhanced irreversibility in
isothermal $M(H)$ or, equivalently, the enhanced critical current
density ($J_c$) at high fields apart from the central peak which
occurs around zero field. In type-II superconductors, the magnetic
fields above the lower critical field ($H_{c1}$) penetrate the
bulk of the sample in the form of flux lines or vortices.
Different mechanisms based on the vortex dynamics have been
discussed to explain the SP in high-$T_C$ cuprates which include
inhomogeneity of the sample,\cite{daeumling,klein} matching
effect,\cite{yoshizaki} surface barriers,\cite{kopylov}
geometrical effects,\cite{bugo} dynamic effects,\cite{elbaum}
structural phase transition in the vortex lattice
(VL),\cite{rosenstein} vortex order-disorder phase
transition,\cite{khaykovich,ertas,giller,radzyner} crossover from
elastic to plastic creep,\cite{abulafia} etc. However, in spite of
plenty of studies dealing with this phenomenon, the general
understanding lacks a converging trend and the proposed models
appear to be more sample specific.

The recent discovery of superconductivity (SC) in Fe-based
pnictides\cite{kamihara} has renewed the interest in vortex
dynamics. Similar to cuprates, pnictides are also layer-based
superconductors, and exhibit a high $T_C$ and type-II nature. In
contrast, pnictides have less anisotropy and larger coherence
length ($\xi$), thus raising question how these influence the
vortex dynamics in these materials. The appearance of a SP in
$M(H)$ is not an universal phenomenon in different families of
pnictides. For the `122' family (AFe$_2$As$_2$, A = Ba, Sr, Ca,
etc.) the appearance of a SP is sensitive enough to the
compositional elements. The pronounced SP has been observed, for
example, in both hole- and electron-doped Ba-122 compounds, which
has been ascribed to various mechanisms,
\cite{prozo-Ba07,shen,sun,eskildsen,kim,inosov-BFCA,kopeli} but it
remains absent in doped Ca-122 compounds.\cite{pramanik}
Similarly, a SP is not consistently seen in the `1111'
(REFeAsO, RE = La, Nd, Ce, Sm, etc.) and in the `11' (FeTe)
families.\cite{yang,bhoi,taen} However, to the best of our
knowledge the vortex dynamics have not been studied in the `111'
family (AFeAs, A = Li, Na, etc) up to now.

Here we study the fishtail effect and the vortex dynamics in
single crystals of LiFeAs which belong to the 111 family of
pnictides. LiFeAs is an oxygen free compound where superconducting
active FeAs layers are separated by Li atoms along the
$c$-axis.\cite{tapp} Remarkably, LiFeAs exhibits SC in absence of
any notable Fermi surface nesting and static
magnetism,\cite{borisenko} however, the presence of
antiferromagnetic fluctuations in the normal
state is inferred from the nuclear magnetic resonance (NMR)
measurements.\cite{jeglic} Interestingly, recent experimental
NMR results\cite{baek} and also theoretical
calculations\cite{hozoi,brydon} indicate a possible $p$-wave SC
state in LiFeAs, which is significantly different from other
families within pnictides. As LiFeAs is nonmagnetic and does not
require any chemical doping to become superconducting, therefore
the FeAs layers are more homogeneous and the crystal is devoid
of coexisting magnetic phases which make it an ideal playground to
study the vortex dynamics. Based on the sharpness of the rocking
curve, recent small angle neutron scattering (SANS) measurements
have revealed that that VL in LiFeAs exhibits no long range order,
however, better ordering than the doped Ba-122
compounds.\cite{inosov,eskildsen,inosov-BFCA}

We have investigated the properties of the vortex state in LiFeAs
by means of isothermal $M(H)$ and magnetic relaxation measurements
which are the most extensively used tools for a variety of
superconducting materials.\cite{yeshurun,blatter} Our results
imply a pronounced SP in both $M(H)$ and $J_c(H)$ at low
temperatures ($T$) with the applied field parallel to the
$c$-axis. We do not find a SP with $H||ab$ plane. We determine the
$J_c$ which is reasonably high, however, lower than those for
doped Ba-122 and 1111 compounds. The magnetic relaxation data, on
the other hand, imply a nonlogarithmic time ($t$) dependence with
the estimated normalized magnetic relaxation rate ($S$) being very
low (even lower than in Ba-122 and 1111 compounds) signifying a
strong pinning for LiFeAs. This specific behavior of the magnetic
relaxation is similar to those of triplet superconductors, i.e.,
Sr$_2$RuO$_4$.\cite{elisabeth} Our data also indicate that the SP in
LiFeAs may be due to a VL phase transition. From the $M(H)$ plots
we have constructed the vortex phase diagram on the $H$-$T$ plane
for LiFeAs.

Single crystals of LiFeAs have been grown using the self-flux
method as detailed in Ref. \onlinecite{igor}. The good quality and
homogeneity of the crystals are confirmed by a exceptionally small
nuclear quadrupole resonance (NQR) linewidth of 64 kHz, a very low
residual resistance of 0.025 m$\Omega$cm and a sharp transition in
specific heat data.\cite{igor,stockert} For the present studies,
two crystals ($S1$ and $S2$) of the same batch with rectangular
shape have been selected. For the magnetic hysteresis loop sample
$S1$ (3.53 $\times$ 2.5 $\times$ 0.21 mm$^3$) and for the magnetic
relaxation measurements sample $S2$ (2.89 $\times$ 2.16 $\times$
0.38 mm$^3$) have been used. Magnetization have been measured
in a Quantum Design MPMS-XL SQUID. Adequate care has
been taken to avoid the exposure of the sample to air before
mounting it in the magnetometer. All the $M(H)$ and $M(t)$
measurements have been performed after cooling the sample in zero
magnetic field from much above $T_C$ to the specific temperature.
The $M(H)$ loops have been investigated with the field up to 50
kOe. For the relaxation measurements, the magnetization has been
measured as function of time for about 8000 s.

\begin{figure}
    \centering
        \includegraphics[width=5.5cm]{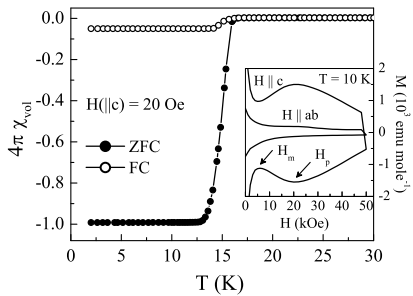}
    \caption{Temperature dependence of $\chi_{vol}$ for LiFeAs deduced from dc-$M$ measured following ZFC and FC protocols. The data have been collected in $H||c$ = 20 Oe. The inset shows $M$ vs $H$ plots measured at 10 K with the field parallel to both $c$-axis and $ab$-plane.}
    \label{fig:Fig1}
\end{figure}

The main panel of Fig. 1 presents the temperature dependence of
the volume susceptibility ($\chi_{vol}$) measured following the
zero-field-cooled (ZFC) and field-cooled (FC) protocols for
magnetization measurements. $\chi_{vol}$ has been deduced from the
measured dc-$M$ with a field of 20 Oe applied parallel to the
crystallographic $c$-axis. The data has been corrected for
demagnetization effects.\cite{osborn} It is evident from Fig. 1
that the sample exhibits bulk SC as characterized by the
diamagnetic signal at low $T$. The sharp transition as well as the
high value of $\chi_{vol}$ in $M_{ZFC}$ demonstrate the high
quality of our crystal. We determine $T_C$ from the bifurcation
point between ZFC and FC branches of the magnetization to be
around 16.5(5) K. In the inset of the Fig. 1 we have plotted the
$M(H)$ data at 10 K with $H$ parallel to both $c$-axis and
$ab$-plane. With increasing $H$ for $H||c$, the magnetic
irreversibility ($M_{irr}$) initially decreases showing a minimum
at a field $H_m$. On further increase in $H$, $M_{irr}$ increases
and exhibits a peak (SP) at a field $H_p$. However, we do not find
any trace of a SP for $H||ab$-plane. While this significant
anisotropic behavior in appearance of the SP is similar to other
pnictide superconductors,\cite{prozo-Ba07} but it remains different
from cuprates, i.e., La$_{1-x}$Sr$_x$CuO$_4$.\cite{radzyner}

\begin{figure}
    \centering
        \includegraphics[width=5.5cm]{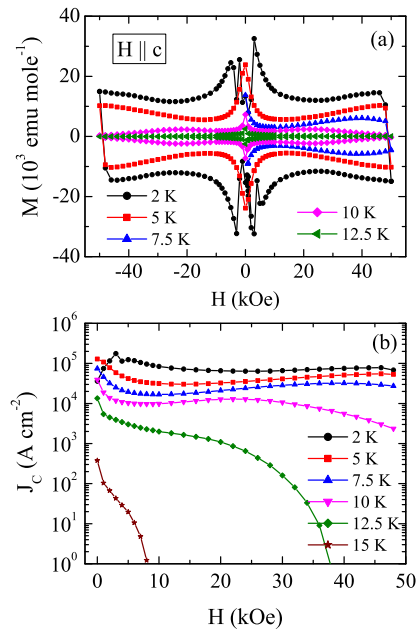}
    \caption{(Color online) (a) The isothermal $M$ vs $H$ loops recorded at different temperatures with $H||c$ for LiFeAs. (b) The critical current $J_c$ as calculated from the $M(H)$ loops in (a) as function of field for different temperatures; for details see text.}
    \label{fig:Fig2}
\end{figure}

Since the SP in $M(H)$ is only evident for $H||c$ in this
compound, we have collected the $M(H||C)$ isotherms at different
$T$ within the SC regime in order to understand the SP
characteristics. The data are plotted in Fig. 2a where $M$ vs $H$
loops are quite symmetric with respect to both the sweeping
direction as well as the polarity of the magnetic field. At low
$T$, however, even though the onset of the SP is evident, the SP
cannot be observed within the measurable field range.
Interestingly, $M(H)$ at 2 K exhibits irregular jumps close to $H$
= 0. These jumps are commonly known as `flux jump'
effects\cite{henry} and will be presented elsewhere in more
detail. With the increase in $T$, a clear SP can be observed in
the $M(H)$ loops. Moreover, we find a field $H_{irr}$ above which
$M_{irr}$ in data vanishes. With increasing $T$, all the
characteristics fields, i.e., $H_m$, $H_p$ and $H_{irr}$ decrease,
and their $T$ variation will be discussed in a later section.

From the magnetic irreversibility in $M(H)$ we have calculated the
critical current $J_c$ exploiting the Bean's critical state
model\cite{bean} $J_c = 20 \Delta M /[a\left(1-a/3b\right)]$,
where $\Delta M$ = $M_{dn}$ - $M_{up}$, $M_{up}$ and $M_{dn}$ are
the magnetization measured with increasing and decreasing field,
respectively, and $a$ and $b$ ($b > a$) are the dimensions of the
crystal perpendicular to the applied $H$. The unit of $\Delta M$
is in emu/cm$^3$, $a$ and $b$ are in cm and the calculated $J_c$
is in A/cm$^2$. The calculated $J_c(H)$ has been plotted in Fig.
2b for different $T$. The variation in $J_c(H)$ is nonmonotonic
and exhibits a broad peak (SP) in the high field region, which is
in line with $\Delta M$ in Fig. 2a. At low $T$, $J_c$ is
rather high, however, its value still being lower than in
doped Ba-122 and 1111
compounds where $J_c$ $\sim$ 10$^{6}$ or even higher.\cite{prozo-Ba07,shen,sun,kim,bhoi} This is in agreement
with the level of disorder as revealed from the SANS
studies.\cite{inosov,eskildsen,inosov-BFCA}

\begin{figure}
    \centering
        \includegraphics[width=6.5cm]{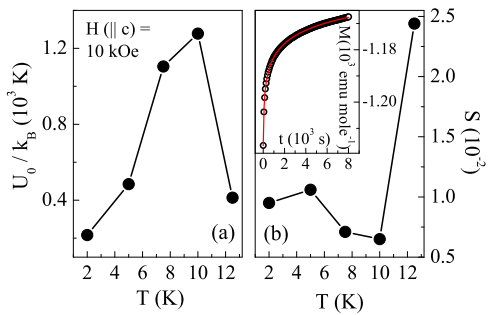}
    \caption{(Color online) (a) The energy barrier $U_0/k_B$ obtained by fitting of Eq. 1 as a function of temperature. (b) The magnetic relaxation rate $S$ as function of temperature in $H||c$ = 10 kOe as calculated using Eq. 2 at $t$ = 1000 s. The inset exemplary shows the best fit of the magnetic relaxation data at $T$ = 10 K and $H||c$ = 10 kOe using Eq. 1.}
    \label{fig:Fig3}
\end{figure}

To understand the origin of the SP and the reasonably high
$J_c$-values in LiFeAs, we have studied detailed vortex dynamics
in this compound by means of $T$- and $H$-dependent magnetic
relaxation measurements.\cite{yeshurun,blatter} Magnetic
relaxation in superconductors is a result of non-equilibrium
spatial arrangement of vortices due to pinning sites. External
applied magnetic field exerts Lorentz forces on the vortices
resulting in their movement, which causes a change in $M(t)$. In
contrast to the original Anderson-Kim model,\cite{anderson} our
relaxation data exhibit a nonlogarithmic time dependence and can
be best fitted with the interpolation formula:\cite{yeshurun}

\begin{eqnarray}
    M(t) = M_0 \left[1 + \frac{\mu k_{\rm B}T}{U_0} \ln
    \left(\frac{t}{t_0}\right)\right]^{-1/\mu},
\end{eqnarray}

where $k_{B}$ is the Boltzmann constant, $U_0$ is the energy
barrier height in absence of a driving force, $t_0$ is the
characteristic relaxation time (usually $\sim 10^{-6} s$ for
type-II superconductors), and $\mu$ is the field-temperature
dependent parameter. This formula yields the normalized magnetic
relaxation rate $S [=(1/M)dM/d\ln(t)]$ as:\cite{yeshurun}

\begin{eqnarray}
    S(t) = \frac{k_{\rm B}T}{U_0 + \mu k_{\rm B}T \ln(t/t_0)}.
\end{eqnarray}

Magnetic relaxation has been measured at different $T$ in $H||c$ =
10 kOe. We find a very slow relaxation, i.e., there is only a 4\%
change in magnetic moment at $H$ = 10 kOe and $T$ = 10 K within a
time period of 8000 s which is much lower than that observed in
cuprates and 122-pnictides.\cite{prozo-Ba07,yeshurun} $U_0/k_B
(T)$ and $S(T)$ as extracted from the fitting of the data
exploiting Eqs. 1 and 2 are plotted in Fig. 3a and 3b,
respectively. One representative fitting of our data has been
included as an inset in Fig. 3b. Surprisingly, we find a
very high value of $t_0$ of the order of 10 s,
which is orders of magnitude higher than for other families in
pnictides and cuprates.\cite{prozo-Ba07,pramanik,yeshurun,
blatter} For LiFeAs the value for the energy barrier $U_0/k_B$
increases with $T$, however, above 10 K it drops down due to
larger thermal fluctuations. A similar behavior can be detected for
$S(T)$, which also shows a non-monotonous behavior (calculated at
$t$ = 1000 s) with a steep decrease followed by an
increase upon lowering temperature.

\begin{figure}[t]
    \centering
        \includegraphics[width=7.5cm]{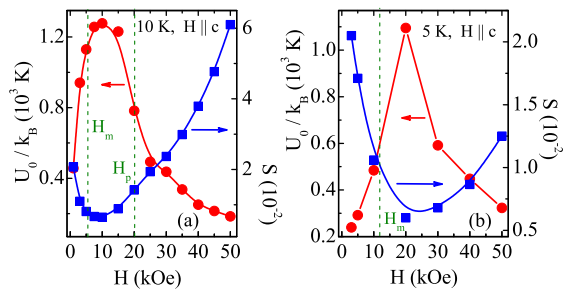}
    \caption{(Color online) (a) The field dependence of $U_0/k_B$ and $S$ extracted from the magnetic relaxation data at 10 K exploiting Eqs. 1 and 2. The vertical dashed lines mark $H_m$ and $H_p$; details see text. (b) The same physical properties $U_0/k_B$ and $S$ shown for 5 K.}
    \label{fig:Fig4}
\end{figure}

In addition to the temperature dependent relaxation studies,
$M(t)$ has been measured in different fields ($H||c$) along the
hysteresis loop at constant $T$. Similarly, we have extracted
$U_0/k_B$ and $S$ using Eqs. 1 and 2. Fig. 4a presents the results
for $U_0/k_B(H)$ and $S(H)$ at 10 K. Here, the vertical dashed
lines in Fig. 4a represent the minimum and maximum (SP) in $M(H)$
at 10 K (see Fig. 1 and 2). It is evident from the plot that
$U_0/k_B$ initially increases with applied fields and then
decreases, showing a peak at a field in between $H_m$ and $H_p$,
where the related calculated $S(H)$ exhibits a minimum. This
variation in $S(H)$ is similar to doped Ba-122
compounds.\cite{prozo-Ba07,kopeli} To examine the behavior of
$U_0/k_B(H)$ and $S(H)$ at low-$T$ where the SP shifts
significantly to higher field values, we have calculated the
parameters at 5 K following the same method. Our results show that
although the field, where $U_0/k_B$ ($S$) exhibits a maximum
(minimum), increases from $\approx$10 kOe at 10 K to $\approx$20
kOe at 5 K (Fig. 4b), this variation does not scale with the large
increase in $H_p$ at low $T$ (see Fig. 2a). Note, that the
calculated $S$ in Figs. 3 and 4 is very low, and that the values
even go below those for Ba-122 and 1111 compounds, where $S >$
0.01.\cite{prozo-Ba07,kopeli,yang,bhoi} The low value of $S$
paired with our very high $t_0$-value imply a high pinning in
LiFeAs. This is quite intriguing as $J_c$ in LiFeAs is lower than
in Ba-122 and 1111 materials,\cite{prozo-Ba07,shen,sun,kim, bhoi}
which means that other type of pinnings rather than the
conventional defects are active in LiFeAs. Indeed, our crystal is
of good quality and homogeneous as discussed earlier. However,
such slow magnetic relaxation has been observed in the $p$-type
superconductor Sr$_2$RuO$_4$ ($S$ $\sim$ 10$^{-3}$), where it has
been pointed out that the observed strong pinning is related to
the superconducting phase of triplet nature rather than conventional
defects.\cite{elisabeth} This raises the question if the observed
low $S$-value in combination with the observed $J_c$ in our single
crystal of LiFeAs might be
within the lines of other studies, which recently proposed triplet
pairing in LiFeAs,\cite{baek,hozoi,brydon} but this requires
further detailed investigations in the future.

From the so far obtained data we have constructed the vortex phase
diagram on the $H$-$T$ plane for LiFeAs (Fig. 5). Above $H_{irr}$,
vortices are in a liquid state. Below $H_{irr}$, they are in a
solid state, however, its nature changes between different field
regimes indicated in the figure (named I, II and III). All the
characteristic fields show a strong $T$ dependence where the data
can be fitted well with the functional form $H_x(T) = H_x(0)(1 -
T/T_C)^n$. We obtain $H_{irr}(0)$ = 299.5(9) kOe and $n$ =
1.46(4), $H_{p}(0)$ = 105.7(8) kOe and $n$ = 1.66(3), $H_{m}(0)$ =
29.1(4) kOe and $n$ = 1.56(3). The values of the exponent $n$ are
reasonably consistent with those for other
pnictides.\cite{prozo-Ba07,yang,shen,bhoi}

\begin{figure}
    \centering
        \includegraphics[width=5.5cm]{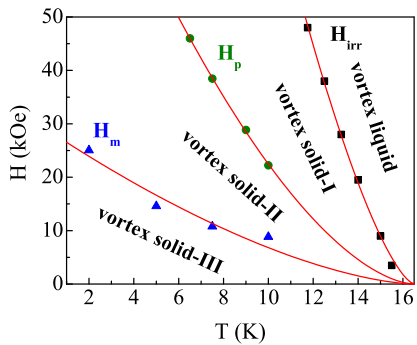}
    \caption{(Color online) The vortex phase diagram on the $H$-$T$ plane with $H||c$ for LiFeAs. Solid lines represent fits of the data using the functional form $H_x(T) = H_x(0)(1 - T/T_C)^n$ (see text).}
    \label{fig:Fig5}
\end{figure}

Now we discuss the origin of the SP effect and the intriguing
vortex dynamics in LiFeAs. The strong $T$ dependence of the
transition lines (Fig. 5) discards the possibility that the SP in
LiFeAs arises due to a vortex order-disorder phase
transition.\cite{khaykovich,ertas,giller,radzyner} Moreover, an
absence of 'mirror-image' correlation between $M(H)$ and $S(H)$ in
Figs. 2 and 4 imply that the dynamic model\cite{elbaum} is not
valid in the present case. Although, the nature of $U_0/k_B(H)$ in
Fig. 4 has qualitative similarity to the model which predicts the
SP being associated with a crossover in flux dynamics from elastic
($<H_p$) to plastic ($>H_p$) creep with increasing
field,\cite{abulafia} the peak in $U_0/k_B(H)$ for LiFeAs occurs
much below the SP, and in low-$T$ at 5 K this mismatch increases.
Moreover, this model predicts $H_p \propto [1 -
(T/T_C)^{4}]^{1.4}$ which does not describe the present $H_{p}(T)$
dependence shown in Fig. 5. Henceforth, the validity of this
model for LiFeAs is questionable.

On the other hand, a structural phase transition in the VL is also
an attractive and possible model which argues that the SP is
associated with the transformation of a hexagonal VL to a square
one with field.\cite{rosenstein} The square structure is supported
by the fourfold symmetry of the intervortex interaction which can
originate in various situations, like, for the anisotropic
($d$-wave) nature of the superconducting gap as in
La$_{2-x}$Sr$_{x}$CuO$_4$,\cite{gilardi} for materials with low
Ginzburg-Landau (GL) parameters ($\kappa$) as in
YNi$_2$B$_2$C,\cite{paul} in the extended GL theories with more
than one order parameter as for the $p$-wave superconductor
Sr$_2$RuO$_4$,\cite{riseman} etc. Recently, such scenario of a
structural phase transition in the VL has been proposed in
Ba(Fe$_{0.925}$Co$_{0.075}$)$_2$As$_2$ where a minimum in $S(T)$
and $S(H)$ has been found.\cite{kopeli} Altogether, taking into
account a similar behavior with a minimum in $S(T)$ and $S(H)$ in
LiFeAs, paired with a comparatively low value of $\kappa$
($\approx$30)\cite{inosov} and the proposed $p$-wave
SC\cite{baek,hozoi,brydon} in LiFeAs, a structural phase
transition in the VL is a possible scenario for the existence of
the SP in this compound. However, further investigations including
microscopic probes are required to confirm these observations.

In summary, single crystalline LiFeAs exhibits a SP in the $M(H)$
loop with $H||c$-axis. The calculated $J_c$s are reasonably high,
however, the values are lower than in the doped Ba-122 and 1111
compounds. We find an extraordinary slow magnetic relaxation
implying a strong pinning which appears not to be related to
conventional defects. Instead, the behavior of magnetic relaxation
is similar to the $p$-wave superconductor Sr$_2$RuO$_4$. We have
constructed the vortex phase diagram on the $H$-$T$ plane for
LiFeAs, with the characterized fields $H_{irr}$, $H_{p}$ and
$H_{m}$ showing a strong $T$ dependence. In accordance with recent
investigations on Ba(Fe$_{0.925}$Co$_{0.075}$)$_2$As$_2$ our data
imply that the SP in LiFeAs most likely originates from a VL phase
transition. Nonetheless, further studies involving microscopic
probes are required to comprehend the SP and vortex dynamics in
this compound.

This work has been supported by the DFG through SPP 1458 and Grant
No. Be1749/13 and WO1532/1-1.

\end{document}